\documentclass[aip,amsmath,amssymb,reprint,]{revtex4-1}

\usepackage{graphicx}
\usepackage{dcolumn}
\usepackage{bm}
\usepackage[utf8]{inputenc}
\usepackage[T1]{fontenc}
\usepackage{mathptmx}
\usepackage{booktabs}
\usepackage{ulem}
\usepackage{xcolor}

\begin{document}

\preprint{AIP/123-QED}

\title{Dynamics of Colloidal Cubes and Cuboids in Cylindrical Nanopores}
\author{Alessandro Patti}

 \email{alessandro.patti@manchester.ac.uk}
 \affiliation{Department of Chemical Engineering and Analytical Science, The University of Manchester, Manchester M13 9PL, United Kingdom}

 \author{Alejandro Cuetos}%
\affiliation{Department of Physical, Chemical and Natural Systems, Pablo de Olavide University, 41013 Seville, Spain}%

\date{\today}

\begin{abstract}
Understanding how colloidal suspensions behave in confined environments has a striking relevance in practical applications. Despite the fact that the behaviour of colloids in the bulk is key to identify the main elements affecting their equilibrium and dynamics, it is only by studying their response under confinement that one can ponder the use of colloids in formulation technology. In particular, confining fluids of anisotropic particles in nanopores provides the opportunity to control their phase behaviour and stabilise a spectrum of morphologies that cannot form in the bulk. By properly selecting pore geometry, particle architecture and system packing, it is possible to tune thermodynamic, structural and dynamical properties for \textit{ad hoc} applications. In the present contribution, we report Grand Canonical and Dynamic Monte Carlo simulations of suspensions of colloidal cubes and cuboids constrained into cylindrical nanopores of different size. We first study their phase behaviour, calculate the chemical potential \textit{vs} density equation of state and characterise the effect of the pore walls on particle anchoring and layering. In particular, at large enough concentrations, we observe the formation of concentric nematic-like coronas of oblate or prolate particles surrounding an isotropic core, whose features resemble those typically detected in the bulk. We then analyse the main characteristics of their dynamics and discover that these are dramatically determined by the ability of particles to diffuse in the longitudinal and radial direction of the nanopore.
\end{abstract}

\maketitle


\section{INTRODUCTION}

The study of the behaviour of simple and complex fluids constrained within nanoscale environments is referred to as nanofluidics. Fluids confined in such small media, which for instance include synthetic nanowires as well as natural nanoporous networks, exhibit thermodynamic, structural and dynamical properties that deviate from those measured in the bulk. In case of nanopore-confined colloids, these deviations are especially significant when the size of the suspended particles or macromolecules is comparable to the size of the hosting nanopore. Experimental \cite{Pieranski1980, Neser1997, cohen2004, Cortes2016, pedrero2021}, theoretical \cite{ponce2001, yao2018, Mizani2019, yao2020, Behzadi2021, teixeira2021} and simulation \cite{avendano2013, patti2013, Muangnapoh2014, workineh2016, avendano2018, luo2019, garlea2019, Anquetil-Deck2020, anzivino2021, donaldson2021} studies on confined colloidal suspensions have shown the existence of a rich realm of nanostructures that are not observed in the bulk. Additionally, depending on the adhesive interactions established between the fluid and the pore walls and their relevance with respect to the cohesive interactions between fluid particles, several new phenomena, such as wetting, capillary condensation and evaporation, layering, anchoring, are detected \cite{dijkstra2001, dijkstra2004, martinez2007, brumby2017, Salgado2019, Basurto2020}. If inter-particle and wall-particle interactions are determined by mere excluded-volume effects, with no enthalpy-driven forces being involved, the behaviour of the confined fluid is exclusively determined by geometric details, which include pore and particle architecture as well as relative particle-to-pore size. 

Due to their ability to form nanostructures that are especially relevant in the design of nanomaterials, anisotropic particles are receiving more and more attention, especially since new accurate experimental techniques for their synthesis have become available \cite{sun2002, Manoharan2003, shankar2004}. In particular, cuboids are especially interesting particles that can self-assemble into a wide spectrum of liquid crystal (LC) phases \cite{john2005, john2008, cuetos2017, patti2018, yang2018}, including the elusive biaxial nematic phase when some conditions (\textit{e.g.} size dispersity, extreme particle anisotropy, application of an external field) are met \cite{belli2011, belli2012, reinink2014, dussi2018, cuetos2019, rafael2021}. Under spherical confinement, experiments and simulations showed that cubes can form face-centered-cubic, hexagonal close-packed or simple-cubic phases depending on the degree of the particle sharpness \cite{wang2018}. By applying Onsager second-order virial theory, Velasco and co-workers studied the effect of constraining perfectly aligned hard board-like particles (HBPs) within two parallel planar hard walls and observed suppression of the nematic-to-smectic transition, usually observed in the bulk, and the occurrence of biaxial phases \cite{varga2010}. A similar study was performed by Mizani \textit{et al}, who applied Parsons-Lee density functional theory within the restricted-orientation Zwanzig model to study the orientational ordering and layering of HBPs in slit pores \cite{Mizani2019}. These authors observed that the hard wall-particle interaction promotes the occurrence of uniaxial nematic order with strong adsorption at the walls. The behaviour of freely-rotating hard cubes confined in between parallel hard walls has been studied by Monte Carlo simulations \cite{khadilkar2016}. In this case, the authors varied the distance between the walls to accommodate between one and five layers of particles, in order to assess what phase transitions occur as the system goes from a quasi-2D geometry to a quasi-3D bulk behavior. Studies of cubes or cuboids under cylindrical confinement are significantly less common. Aggregation of silver cubes in cylindrical nanopores of alumina membranes has been investigated experimentally to establish the effect of the substrate on the cubes plasmonic sensitivity, which was found to be more than one order of magnitude larger than that observed on planar solid substrates \cite{konig2014}. Very recent experiments have assessed the impact of cylindrical confinement on the ability of cellulose nanocrystals (CNCs), whose rod-like shape reminds that of the prolate HBPs studied here, of forming cholesteric (Ch) liquid crystals \cite{prince2021}. The authors observed that nanoconfined CNCs organise into a core-shell structure, with an isotropic core, running parallel to the longitudinal axis of the cylindrical capillary, surrounded by a Ch shell. We will see that similar core-shell morphologies have also been observed in our simulations.

The purpose of the present article is studying the behaviour of colloidal cuboids confined in cylindrical nanopores. More specifically, we investigate equilibrium, structural and dynamical properties of perfect hard cubes as well as oblate and prolate HBPs constrained in cylindrical nanopores of different diameters. There are two main reasons why we opted for hard-core potentials. To the best of our knowledge, there are no available soft potentials that could be employed to efficiently estimate the minimum distance between pairs of cubes or cuboids, and the qualitative insight that underpins the physics of these systems would not be significantly affected by the specific choice of the repulsive potential. This has been observed by comparing the phase behaviour of hard spherocylinders \cite{BOL97} with soft repulsive oblate \cite{CUE08, MAR09} and prolate \cite{CUE15} spherocylinders or with spherocylindrical particles with Kihara attractive interactions \cite{CUE03}. To determine the system equilibrium properties, we perform Grand Canonical Monte Carlo (GCMC) simulations, whereas the study of the dynamics is carried out by Dynamic Monte Carlo (DMC) simulations. We show that the orientation of particles within the nanopore is not uniform, but changes with its radial distance from the pore wall, and determines their mobility at both short and long-time scales. The present article is organised as follows. In Section II, we describe the model adopted to mimic colloidal cuboids and the main GCMC and DMC simulation details applied to equilibrate the systems, study their thermodynamic and structural properties and produce their time trajectories. Results, which include the chemical potential \textit{vs} density equations of state, structural correlation functions and order parameters, mean square displacements and the self-part of the van Hove correlation functions are presented and discussed in Section III. Finally, in Section IV, we draw our conclusions.


\section{METHODS}

\subsection{Model}
Colloidal cuboids are modelled as HBPs of thickness $T$, length $L$, and width $W$. For the particular case of hard cubes (HCs), $T=W=L$. We have set the thickness $T_p$ of prolate HPBs as the system unit length. The dimensions of HCs and HBPs have been set in such a way that the particle volume, $v_0=LWT$, is the same for the three geometries. Details are given in Table \ref{tab:sys_size}. All particles interact via a hard-core potential with each other and with the pore walls. As such, they can freely displace and rotate, with no restrictions on mutual distances and orientations as long as no overlaps are produced. The occurrence of overlaps has been assessed via the  separating axes method described by Gottschalk and coworkers \cite{gottschalk1996} and later adapted by John and Escobedo to determine the phase and aggregation behaviour of tetragonal parallelepipeds \cite{john2005}.

\begin{table}[!h]
    \caption{Reduced length ($L^{\star} \equiv L/T_p$), width ($W^{\star} \equiv W/T_p$), thickness ($T^{\star} \equiv T/T_p$) and volume ($v^{\star}_0 \equiv v_0/T_p^3$) of the three sets of particles studied in this work.}
    \centering
    \begin{ruledtabular}
    \begin{tabular}{lcccc}
         & Perfect Cube \,& Oblate Cuboid  \,& Prolate Cuboid \, \\
        \midrule
        $L^{\star}$   & $\sqrt[3]{9}$ & 3 & 9  \\
        $W^{\star}$   & $\sqrt[3]{9}$ & 3 & 1  \\
        $T^{\star}$   & $\sqrt[3]{9}$ & 1 & 1  \\
        $v^{\star}_0$ & 9             & 9 & 9  \\
    \end{tabular}
    \end{ruledtabular}
    \label{tab:sys_size}
\end{table}


\subsection{Simulation Methodology}
Simulations comprised equilibration and production runs. Equilibration runs consisted of standard MC simulations that were performed in the grand canonical ensemble at constant pore volume, $V$, and chemical potential, $\beta \mu$, with $\beta$ the energy unit \cite{FRENKELSMIT}. We then computed the excess chemical potential from the ensemble density $\rho=\langle N \rangle /V$, with $\langle N \rangle$ the average number of particles in the nanopore, through the equation:

\begin{equation}
\label{eq1}
    \beta\mu^{\rm ex}=\beta\mu-\ln \rho.
\end{equation}

\begin{figure}[!t]
\includegraphics[width=\columnwidth]{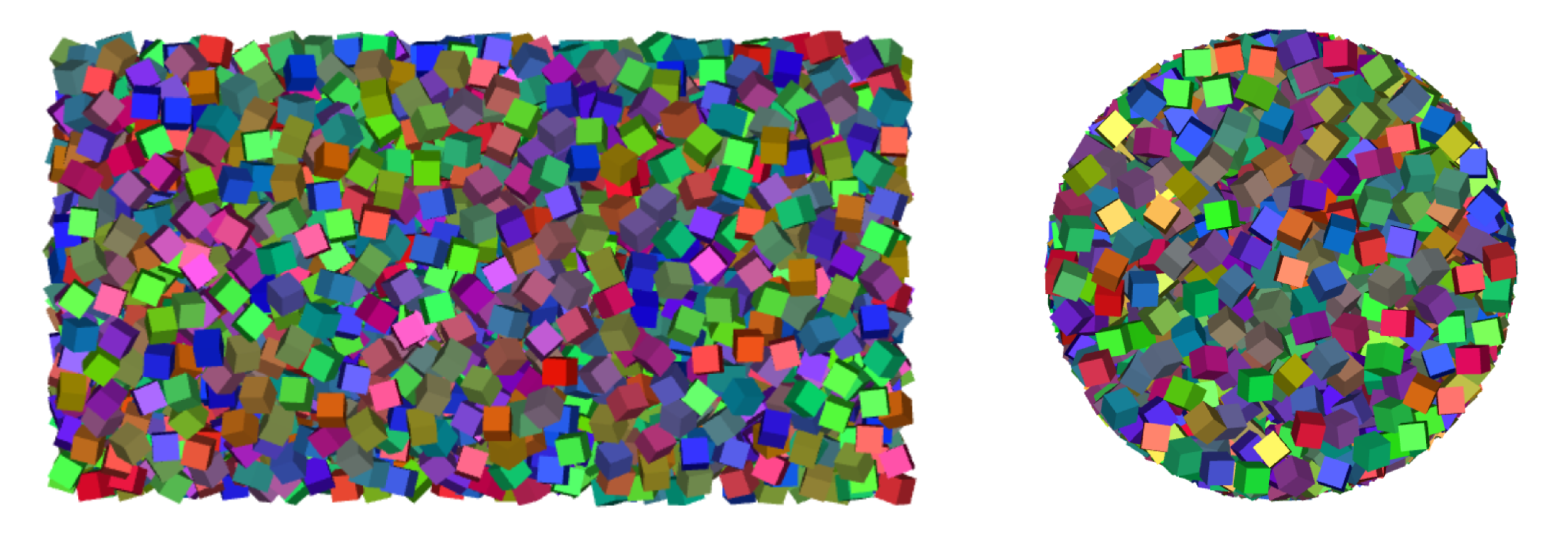}
\caption{Longitudinal (left) and frontal (right) views of an exemplary equilibrium configuration obtained at $\beta\mu^{ex}=5.011$ of systems containing HCs in cylindrical nanopores of radius $R^{\star}=9\times \sqrt[3]{9}$. Colour gradient indicates different particle orientations. In particular, particles in blue have their reference unit vector oriented along the pore longitudinal axis, while particles in red or green have their reference unit vector oriented perpendicularly to this axis.}
\label{fig:config}
\end{figure}

GCMC simulations were also employed to determine the thermodynamic and structural properties of the confined fluids of HCs and HBPs. By contrast, to study their dynamics, DMC simulations were performed in the canonical ensemble at constant $V$ and number of particles, $N$. In all the cases, the particles have been confined in cylindrical nanopores with periodic boundaries in their longitudinal direction. Radius and length of the nanopore have been set depending on the dimensions of the three families of particles studied here. More specifically, HCs have been arranged in nanopores of radius $R^{\star} \equiv R/T_p = \sqrt[3]{9} \times \{3, 6, 9\}$, while oblate and prolate HBPs in nanopores of radius $R^{\star} = \{9, 12, 15\}$ and $R^{\star}= \{12, 15, 22.5\}$, respectively. By contrast, the nanopore length, $h>>L$, has been set to observe a statistically relevant number of particles at the smallest chemical potentials. An exemplary configuration of a nanopore hosting HCs is displayed in Fig.\,\ref{fig:config} for $\beta\mu^{ex}=5.011$. Specific details of equilibration and production runs are provided below. 

\subsubsection{Equilibration}
Initial configurations consisted of randomly positioned, but perfectly aligned HBPs with their main axis oriented along the longitudinal direction of the nanopore. In some cases, equilibrium configurations were employed as the initial configurations of systems at larger chemical potentials. Equilibration was considered accomplished when $N$ or, equivalently, the packing fraction $\eta \equiv Nv_0/V$ achieved a stationary value within reasonable statistical fluctuations. We also computed the total nematic order parameters, which usually take longer to relax as compared to the packing fraction. To this end, we applied the standard procedure of diagonalising the following traceless symmetric second-rank tensor\cite{eppenga1984}

\begin{equation}
\label{eq2}
    {\bf Q}^{\lambda \lambda} = \frac{1}{2N} \Big \langle \sum_{i=1}^N \left( 3 \hat{\lambda}_i \cdot \hat{\lambda}_i - {\bf I} \right) \Big \rangle
\end{equation}

\noindent where $i$ indicates a generic particle, ${\bf I}$ the second-rank unit tensor and $\hat{\lambda}_i = \hat{T}_i$, $\hat{W}_i$, $\hat{L}_i$ the unit vector aligned with the particle thickness, width and length, respectively. The eigenvalues of ${\bf Q^{\lambda \lambda}}$ and their relative eigenvectors correspond, respectively, to the order parameters ($S_{2,W}$, $S_{2,T}$, $S_{2,L}$) and nematic directors ($\hat{m}$, $\hat{p}$, $\hat{n}$) coupled to each particle axis. In the isotropic phase, the order parameters are all close to zero, whereas if particles tend to align, then their preferential orientation is identified by the largest positive eigenvalue and its associated eigenvector. The interested reader is referred to our previous works for additional details \cite{cuetos2017, patti2018, cuetos2019}. 


\subsubsection{Production}
\begin{figure}[h!]
\includegraphics[width=\columnwidth]{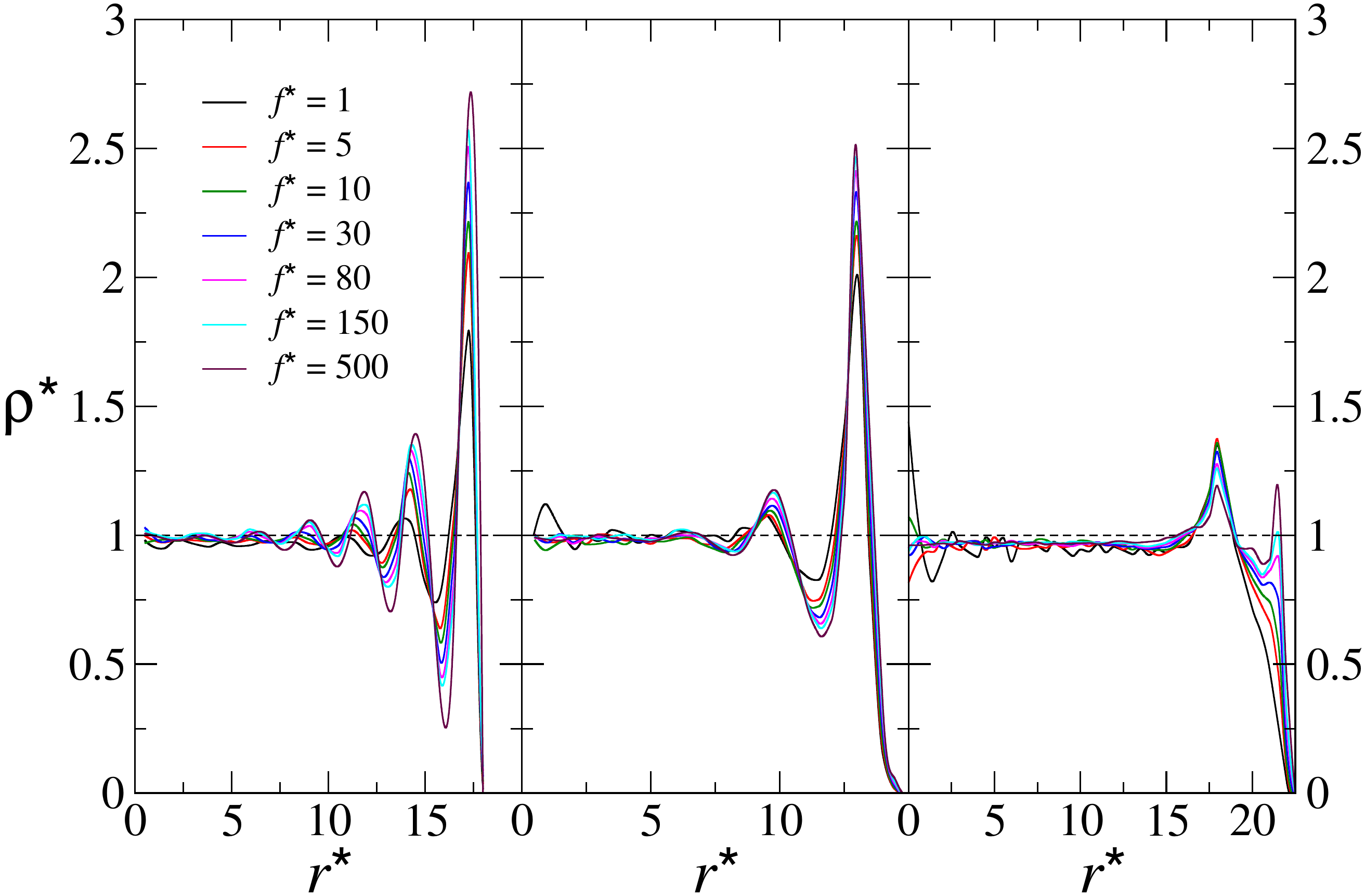}
\caption{Normalised density, $\rho^{\star}\equiv \rho/\rho_{\rm bulk}$, of HCs (left), oblate HBPs (middle) and prolate HBPs (right) as a function of the reduced distance, $r^{\star} \equiv r/T_p$, from the nanopore longitudinal axis. The curves in each of the three frames indicate a different value of the fugacity $f^{\star}=\exp(\beta \mu)$. Nanopore radius from left to right: $R^{\star}=9\times \sqrt[3]{9}$ (HCs), 15 (oblate HBPs), and 22.5 (prolate HBPs).}
\label{fig:dens_prof}
\end{figure}

Once the systems achieved equilibrium, we investigated their structural and dynamical properties by performing, respectively, GCMC and DMC simulations. The initial configurations for the production runs were those equilibrated in the preliminary equilibration step. In particular, in the GCMC production simulations the chemical potential was the same as that imposed in the equilibration run, while in DMC simulations, performed in the canonical ensemble, we set the number density obtained in the equilibration step. To study the impact of confinement, the pore was divided in concentric cylindrical shells. The fluid density in these shells is not homogeneous, but exhibits significant fluctuations in the vicinity of the pore walls. These fluctuations depend on the radial distance from the longitudinal axis, as exemplary shown in Fig.\,\ref{fig:dens_prof}, where we report the numerical density along the nanopore radial direction, normalised with that measured in the bulk at the same chemical potential, for HCs (left frame) and oblate (middle frame) and prolate (right frame) HBPs. We observe that, depending on the geometry of particles and their ability to effectively pack at the pore wall, the density distribution profiles change. In particular, the packing is especially effective for cubes that can best adapt to the pore wall's curvature, but significantly less effective for rod-like HBPs.

The relevance of confinement and density fluctuations on particle orientation has been systematically studied by computing the longitudinal and radial orientational order parameters, which measure, respectively, the average alignment of $\hat{T}$, $\hat{W}$ and $\hat{L}$ along the axis of the nanopore and perpendicularly to it. Both sets of order parameters have been obtained by diagonalisation of the traceless second-rank symmetric tensor that reads

\begin{equation}
\label{eq3}
\textbf{Q}^{\nu \nu}_{\theta}= \Bigg \langle \frac{1}{2N_{\theta}}\sum_{i=1}^{N_{\theta}} \left( 3 \hat{\nu}_i \cdot \hat{\nu}_i - \textbf{I} \right) \Bigg \rangle_{\theta}
\end{equation}

\noindent where $N_{\theta}$ refers to the instantaneous number of particles contained in the generic cylindrical shell $\theta$ at distance $r$ from the nanopore axis, $\hat{\nu}_i$ is the vector projection of $\hat{T}_i$, $\hat{W}_i$ or $\hat{L}_i$ on the radial or longitudinal direction of the cylindrical nanopore, and $\langle ... \rangle_{\theta}$ indicates ensemble average within the cylindrical shell $\theta$. We notice that $\textbf{Q}^{\nu \nu}_{\theta}$ is formally different from  $\textbf{Q}^{\lambda \lambda}$, which has been defined in Eq.\,(2): the former measures how the particles in a given shell orient along the nanopore radial and longitudinal directions, whereas the latter measures the average orientation of all particles in the whole nanopore in the laboratory frame of reference. In particular, the nematic order parameter $S^{\theta}_{2,L_r}$, which measures the alignment of $\hat{L}$ along the radial ($r$) direction of the nanopore in the cylindrical shell $\theta$, is the largest eigenvalue of the tensor $\textbf{Q}^{L_r L_r}_{\theta}$. Similarly, the nematic order parameter $S^{\theta}_{2,L_l}$, which measures the alignment of $\hat{L}$ along the longitudinal ($l$) direction of the nanopore in the cylindrical shell $\theta$ is obtained from the largest eigenvalue of tensor $\textbf{Q}^{L_l L_l}_{\theta}$. From these definitions, it follows that if, in a given cylindrical shell $\theta$, prolate HBPs are perfectly aligned along the nanopore longitudinal axis, then $S^{\theta}_{2,L_l}=1$ and $S^{\theta}_{2,L_r}=-0.5$. By contrast, if they are aligned along the nanopore radial direction, then $S^{\theta}_{2,L_r}=1$ and $S^{\theta}_{2,L_l}=-0.5$. Similar expressions to those valid for $\hat{L}$ have been employed to estimate the radial and longitudinal alignment of $\hat{T}$ and $\hat{W}$. 

To study the dynamics, we performed DMC simulations in the canonical ensemble. The DMC method has been discussed elsewhere \cite{patti2012, cuetos2015, corbett2018, garcia2020, chiappini2020}. Here we only present its essential features and refer the interested reader to these works for details. In a DMC cycle, we attempt $N$ combined displacements and rotations of randomly selected particles, which are accepted or rejected with probability $\text{min}[1, \text{exp}(-\beta \Delta E)]$, where $\Delta E$ is the difference in energy between new and old configuration. Since particles interact via a hard-core potential, only moves producing a geometrical overlap are rejected, while all the others are always accepted. In the latter case, the new position of a particle $i$ is set by summing three elementary displacements along each space dimension: $\delta \textbf{r}_i = X_T \hat{T}_i + X_W \hat{W}_i + X_L \hat{L}_i$. The terms $X_{\alpha}$, with $\alpha = T$, $W$, or $L$, are uniform distributions that depend on the particle translational diffusion coefficients at infinite dilution. In particular:

\begin{equation}\label{eq4}
    |X_{\alpha}| \le \sqrt{2D^{\rm tra}_{\alpha,i} \delta t_{\text{MC}}},
\end{equation}

\noindent where $\delta t_{\text{MC}}$ is a DMC time step. Additionally, reorientations of the particle axes, $\hat{T}_i$, $\hat{W}_i$, $\hat{L}_i$, are attempted by imposing three consecutive rigid rotations around $\hat{L}_i$, $\hat{W}_i$, and $\hat{T}_i$, respectively. Similarly to translations, elementary rotations around are constrained within the limits imposed by the rotational diffusion coefficients around the three particle axes at infinite dilution:

\begin{equation}\label{eq5}
|Y_{\alpha}| \le \sqrt{2 D^{\rm rot}_{\alpha,i} \delta t_{\text{MC}}},     
\end{equation}

\noindent Translational and rotational diffusion coefficients, $D^{\rm tra}_{\alpha}$ and $D^{\rm rot}_{\alpha}$, have been obtained by applying the open-source software HYDRO$^{++}$, which describes particles of generic shape as an array of spherical beads of arbitrary size \cite{carrasco1999, delatorre2007}. The values of translational and rotational diffusion coefficients for the geometries relevant to our study, are reported in Table \ref{tab:diff} in units of $T_p^2 \tau^{-1}$ and $\tau^{-1}$, respectively, with $\tau$ the time unit.

\begin{table}[!h]
 \caption{Infinite-dilution translational and rotational diffusion coefficients of HCs and HBPs studied in this work as obtained from HYDRO$^{++}$ software \cite{carrasco1999, delatorre2007}.}
\centering
\begin{ruledtabular}
\begin{tabular}{lccc}
 & Perfect Cube \,& Oblate Cuboid \,& Prolate Cuboid \, \\
 \midrule
$D^{\rm tra}_T \tau/T^2_p \cdot 10^{-2}$ & 4.237 & 3.352 & 2.628 \\ 
$D^{\rm tra}_W \tau/T^2_p \cdot 10^{-2}$ & 4.237 & 3.970 & 2.628 \\ 
$D^{\rm tra}_L \tau/T^2_p \cdot 10^{-2}$ & 4.237 & 3.970 & 3.624 \\ 
$D^{\rm rot}_T \tau \cdot 10^{-2}$ & 1.935 & 0.989 & 0.215 \\
$D^{\rm rot}_W \tau \cdot 10^{-2}$ & 1.935 & 1.332 & 0.215 \\ 
$D^{\rm rot}_L \tau \cdot 10^{-2}$ & 1.935 & 1.332 & 2.992 \\
\end{tabular}
\end{ruledtabular}
\label{tab:diff}
\end{table}

Regardless the DMC time step employed in Eqs.\,(2) and (3), the actual time scale of Brownian dynamics, $t_{\text{BD}}$, must be recovered \cite{patti2012, cuetos2015, corbett2018, garcia2020, chiappini2020}. The rescaling to the BD timescale is crucial to ensure a consistent comparison between the dynamics of different families of particles (\textit{e.g.} HCs and HBPs) and between the dynamics of the same system under different conditions (\textit{e.g.} HCs at low and high density). To this end, the DMC time scale must be rescaled via the acceptance rate $\mathcal{A}$:

\begin{equation}\label{eq6}
    t_{\rm BD} = \frac{\mathcal{A}}{3} C_{\rm MC} \delta t_{\text{MC}}
\end{equation}

\noindent where $C_{\rm MC}$ is the total number of DMC cycles.

A key aspect that we further stress here is the fluid spatial inhomogeneity and the existence of density fluctuations as shown in Fig.\,\ref{fig:dens_prof}. Because a change in density affects the probability of accepting a particle move, the acceptance rate in Eq.\,(4) is not space invariant, but depends on the radial coordinate. Consequently, following our recent work on the extension of the DMC method to heterogeneous systems \cite{garcia2020}, we have discretised the nanopore into $k$ concentric cylindrical shells, within which $\mathcal{A}$ can be safely assumed constant. From this, it follows that

\begin{equation}\label{eq7}
    3\delta t_{\text{BD}}= \mathcal{A}_1 \delta t_{\text{MC},1}=\mathcal{A}_2 \delta t_{\text{MC},2}= \dots =\mathcal{A}_k \delta t_{\text{MC},k}
\end{equation}

\noindent where $\mathcal{A}_1, \dots, \mathcal{A}_k$ and $\delta t_{\text{MC},1}, \dots, \delta t_{\text{MC},k}$ refer, respectively, to the acceptance rate and MC time step in the cylindrical shell $1,\dots,k$. It follows that Eq.\,\ref{eq7} establishes a link between the DMC time steps across the cylindrical shells and guarantees the existence of a unique BD time scale. In practice,  one of the $k$ shells is arbitrarily selected to set a reference DMC time step, which is kept constant. By contrast, the time steps associated to all the other shells are obtained by convergence of Eq.\,\ref{eq7} over a short preliminary DMC simulation \cite{garcia2020}. In particular, we chose as a reference the farthest shell from the pore wall and, in this shell, we set $\delta t_{\text{MC},1}=10^{-3}\tau$. The so-calculated MC time steps for each shell are then employed in Eqs.\,\ref{eq4} and \ref{eq5} to obtain the maximum displacements and rotations in each cylindrical shell and consistently evaluate and characterise the dynamics of HCs and HBPs in the nanopore. Additional details are available in our recent work \cite{garcia2020}.


\begin{figure}[t!]
\includegraphics[width=\columnwidth]{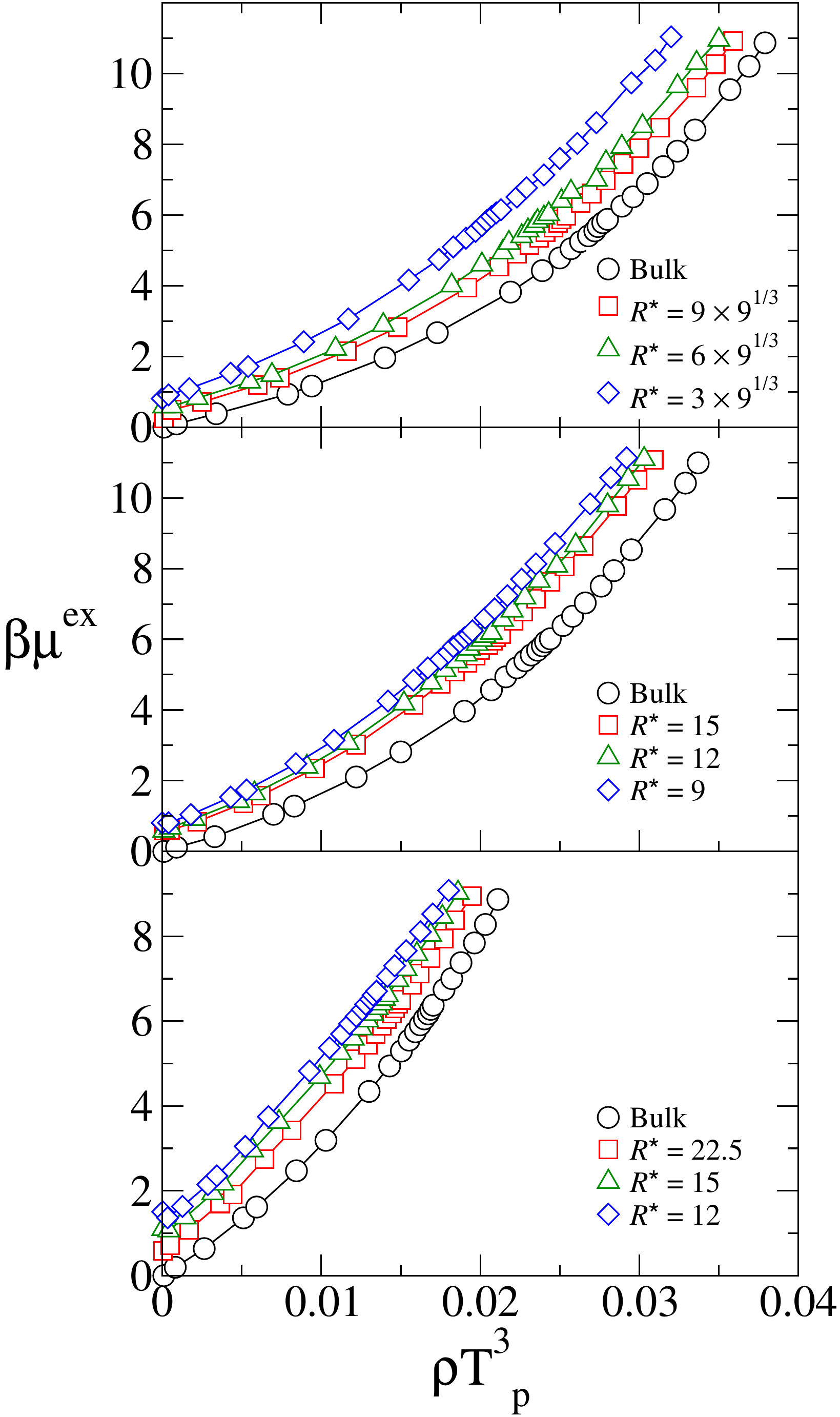}
\caption{Excess chemical potential as a function of the number density in systems of HCs (top), oblate HBPs (middle) and prolate HBPs (bottom). Empty circles refer to systems in the bulk, while squares, triangles and diamonds to systems confined in cylindrical nanopores of different radius. Solid lines are guides for the eye.}
\label{fig:EOS}
\end{figure}

\section{RESULTS AND DISCUSSION}
In this section, we showcase simulation results that highlight the most relevant properties of HCs and HBPs in cylindrical nanopores of different radius. First, we report the impact of confinement on their thermodynamic and structural properties, and then we discuss the dynamics.
 
\begin{figure}[t!]
\includegraphics[width=\columnwidth]{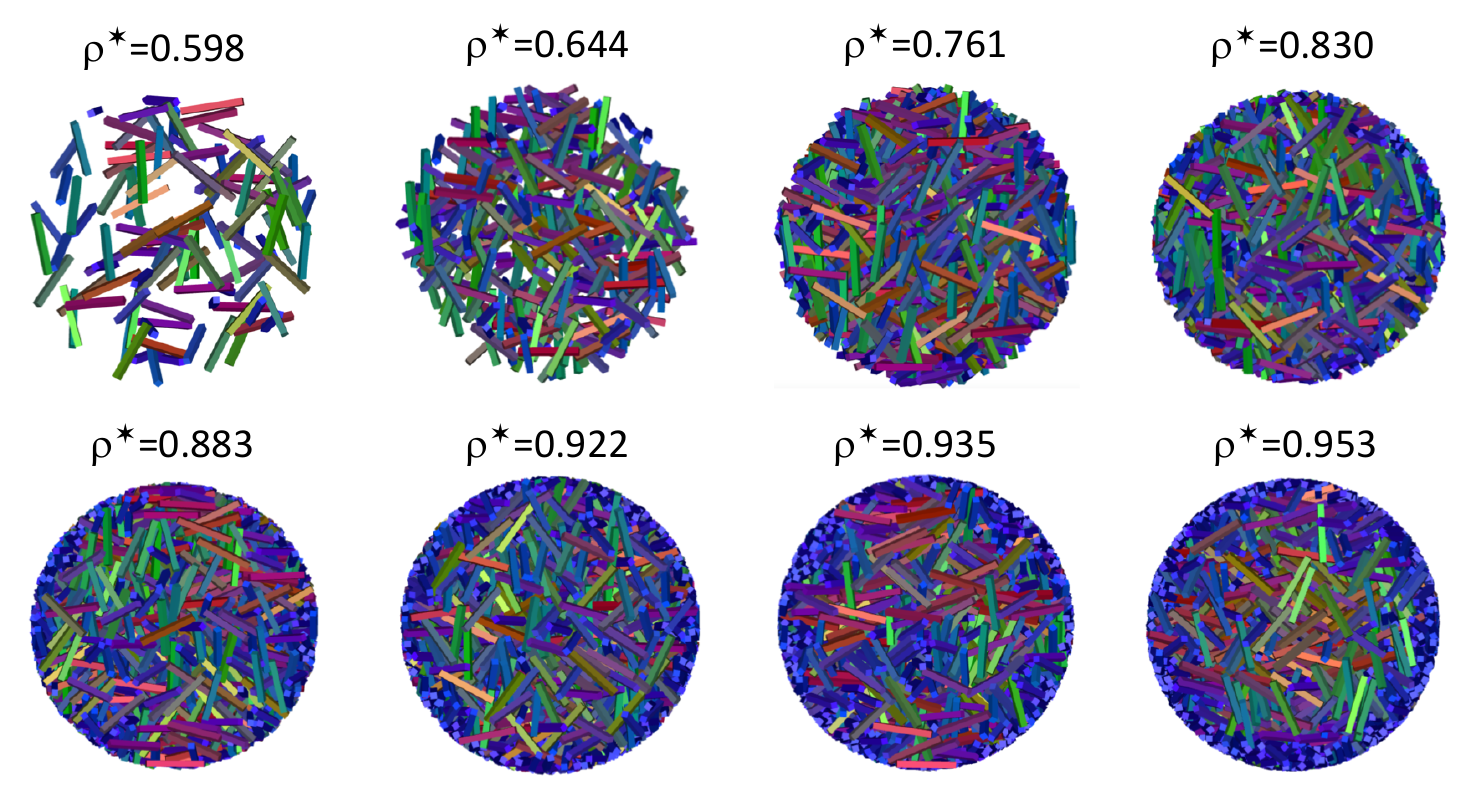}
\caption{Prolate HBPs in cylindrical nanopores of radius $R^{\star}=22.5$ at different normalised densities $\rho^{\star}\equiv \rho/\rho_{\rm bulk}$. The particles are coloured according to the orientation of their major axis. In particular, particles in blue are oriented along the pore longitudinal axis, while particles in red or green are oriented perpendicularly to this axis.}
\label{fig:snaps_prolate}
\end{figure}

In Fig.\,\ref{fig:EOS}, plots of the excess chemical potential \textit{vs} the reduced density are shown for HCs and oblate and prolate HBPs. Each of these three sets of equations of state has been obtained in the bulk and in cylindrical nanopores, as indicated in the figure legends. As a general tendency, the dependence of $\beta \mu^{\rm ex}$ on $\rho$ observed in confined systems deviates significantly from that in the bulk, particularly so at large densities and in narrow tubes, where the effect of walls becomes especially relevant. As an example of the effect of the walls on the confined particles, in Fig.\,\ref{fig:snaps_prolate} we show a sequence of snapshots of prolate HBPs in nanopores of radius $R^{\star}=22.5$ at increasing value of the excess chemical potential and volume fraction. In dilute systems, HBPs appear to be randomly oriented, regardless their position in the pore. At larger densities, we notice a gradually increasing alignment of the particles located close to the walls, whereas those in the centre persist in their disordered state. The alignment along the nanopore longitudinal direction basically sparks the formation of a nematic-like cylindrical layer of HBPs that separates the walls from the disordered inner core. We notice that such nematic-like layers develop at densities that, in the bulk, would be insufficient for the formation of the nematic phase\cite{cuetos2017}, found at $\rho T_p^3 \approx 0.033$, being significantly larger than that of the systems studied here (see bottom frame of Fig.\,\ref{fig:EOS}).

\begin{figure}[t!]
\includegraphics[width=\columnwidth]{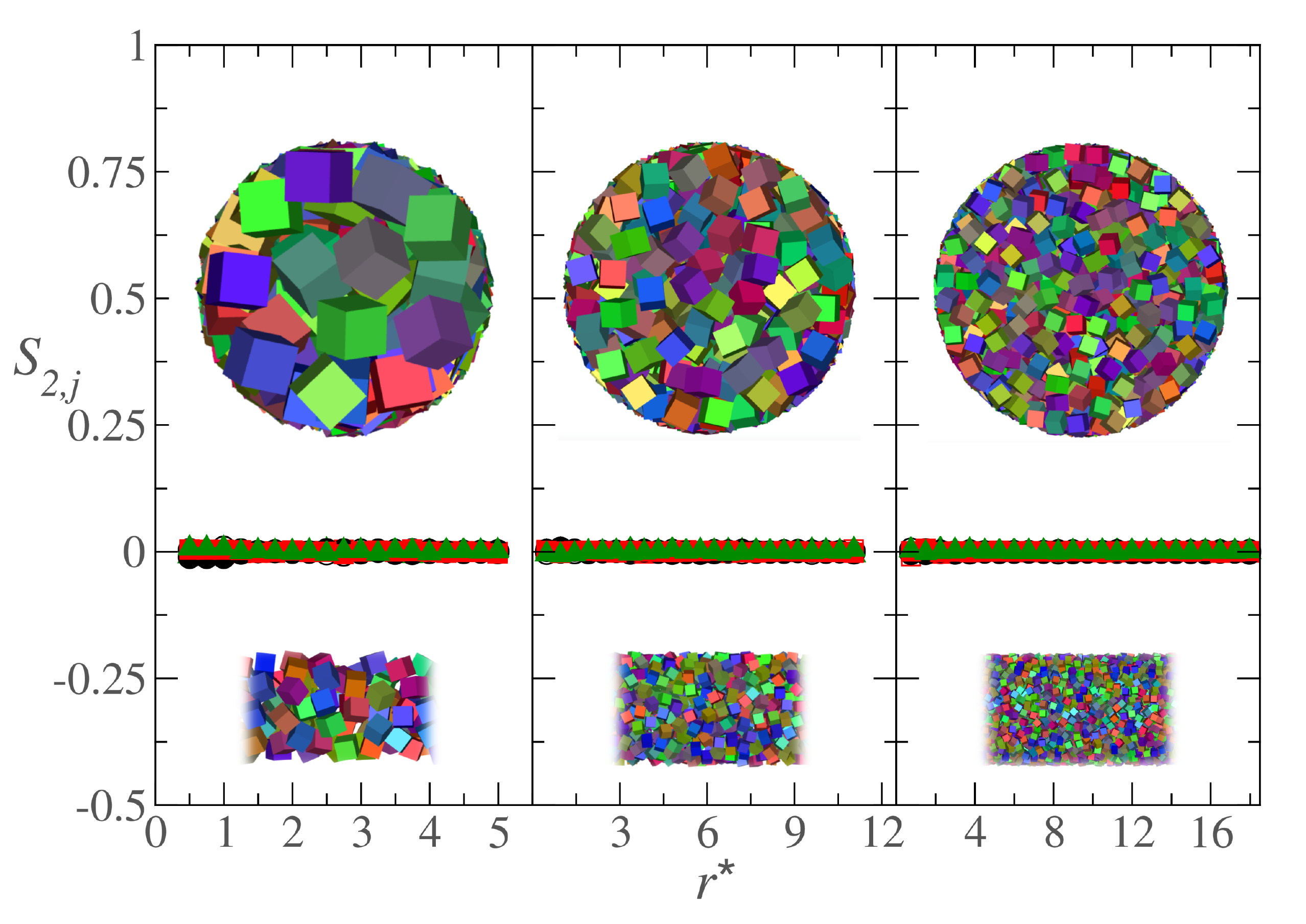}
\caption{Radial ($S^{\theta}_{2,\lambda_r}$, empty symbols) and longitudinal ($S^{\theta}_{2,\lambda_l}$, solid symbols) order parameters of HCs as a function of the distance from the centre of the cylindrical nanopore at $f^{\star}=2000$ and $R^{\star}=3 \times \sqrt[3]{9}$ (left), $R^{\star}=6 \times \sqrt[3]{9}$ (middle) and $R^{\star}=9 \times \sqrt[3]{9}$ (right). Circles, squares and triangles refer, respectively, to the projections of the particle unit vectors aligned with $\hat{\lambda}=\hat{L}$, $\hat{W}$ or $\hat{T}$. The insets in each frame show a frontal and side view of the  nanopore, with the particles coloured according to the orientation of their major axisIn particular, particles in blue are oriented along the pore longitudinal axis, while particles in red or green are oriented perpendicularly to this axis. Particles in different snapshots are not in scale relative to one another. Solid lines are guides for the eye. }
\label{fig:order_cubes}
\end{figure}

\begin{figure}[!t]
\includegraphics[width=\columnwidth]{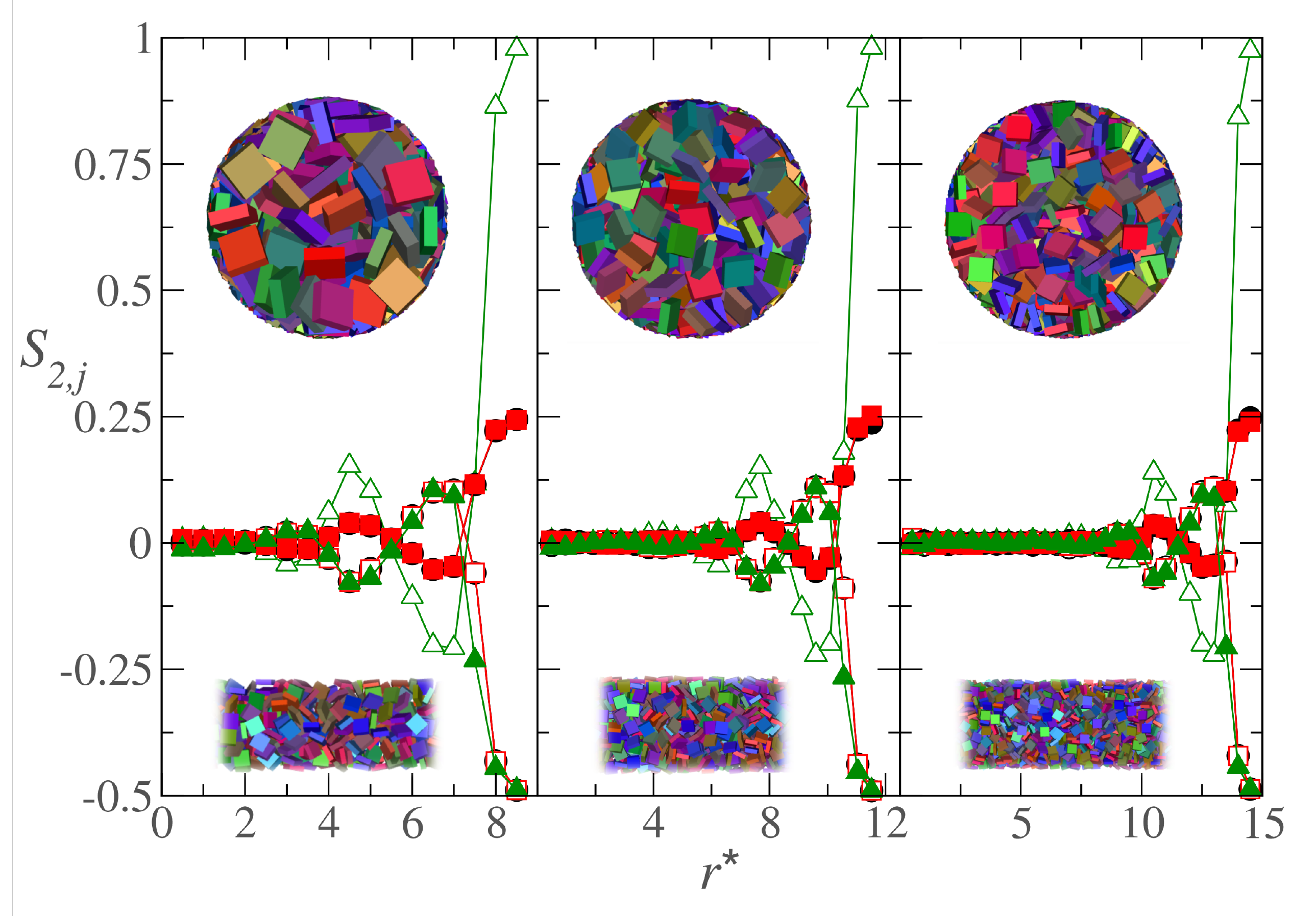}
\caption{Radial ($S^{\theta}_{2,\lambda_r}$, empty symbols) and longitudinal ($S^{\theta}_{2,\lambda_l}$, solid symbols) order parameters of oblate HBPs as a function of the distance from the centre of the cylindrical nanopore at $f^{\star}=2000$ and $R^{\star}=9$ (left), $R^{\star}=12$ (middle) and $R^{\star}=15$ (right). Circles, squares and triangles refer, respectively, to the projections of the particle unit vectors aligned with $\hat{\lambda}=\hat{L}$, $\hat{W}$ or $\hat{T}$. The insets in each frame show a frontal and side view of the nanopore, with the particles coloured according to the orientation of their major axis. Particles in different snapshots are not in scale relative to one another. Solid lines are guides for the eye.}
\label{fig:ob_order_cubes}
\end{figure}

\begin{figure}[!t]
\includegraphics[width=\columnwidth]{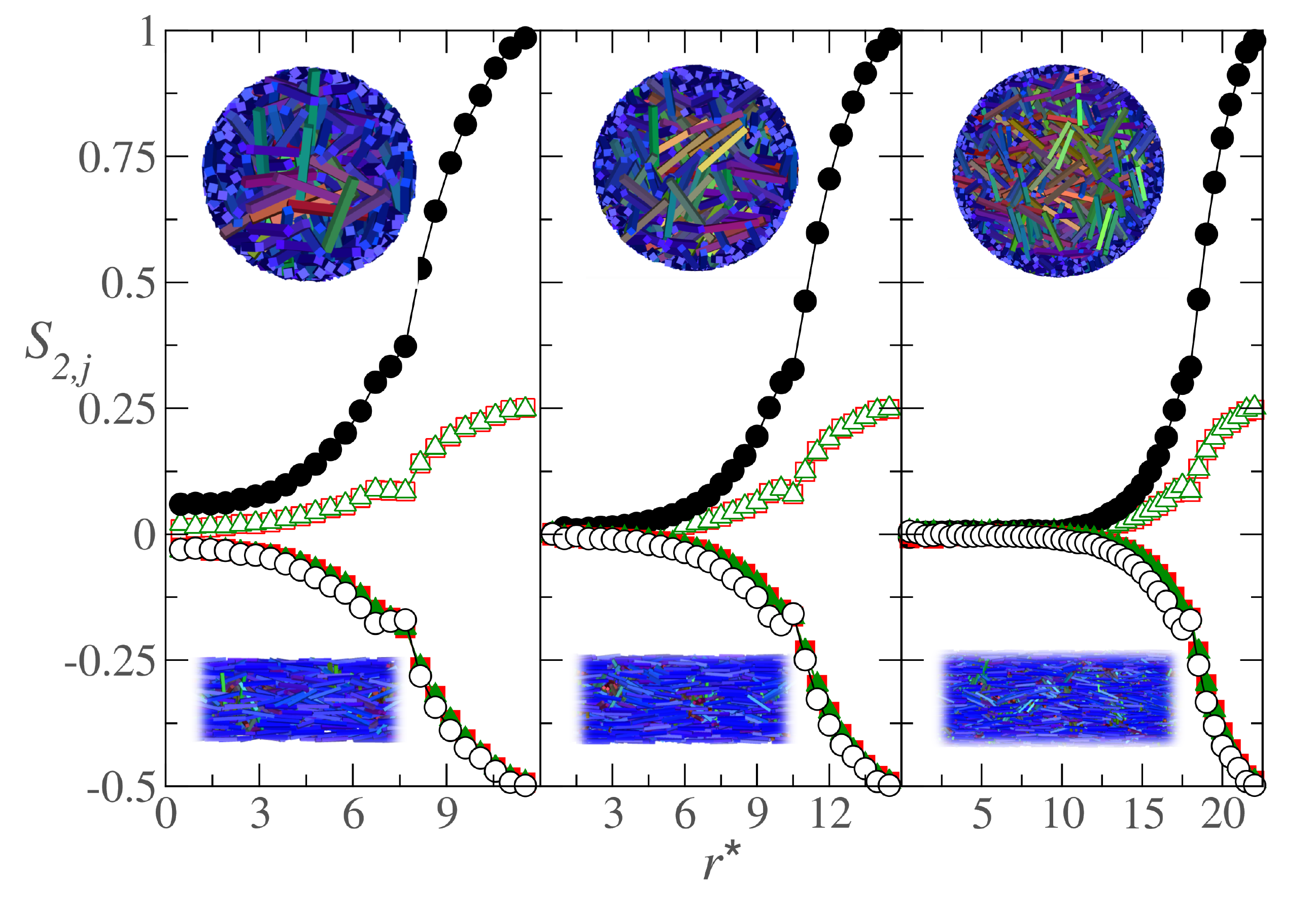}
\caption{Radial ($S^{\theta}_{2,\lambda_r}$, empty symbols) and longitudinal ($S^{\theta}_{2,\lambda_l}$, solid symbols) order parameters of prolate HBPs as a function of the distance from the centre of the cylindrical nanopore at $f^{\star}=5000$ and $R^{\star}=12$ (left), $R^{\star}=15$ (middle) and $R^{\star}=22.5$ (right). Circles, squares and triangles refer, respectively, to the projections of the particle unit vectors aligned with $\hat{\lambda}=\hat{L}$, $\hat{W}$ or $\hat{T}$. The insets in each frame show a frontal and side view of the nanopore, with the particles coloured according to the orientation of their major axis. Particles in different snapshots are not in scale relative to one another. Solid lines are guides for the eye.}
\label{fig:pr_order_cubes}
\end{figure}

The resulting radial and longitudinal nematic order parameters are displayed in Figs.\,\ref{fig:order_cubes}, \ref{fig:ob_order_cubes} and \ref{fig:pr_order_cubes} for HCs, oblate HBPs and prolate HBPs, respectively, at three different nanopore radii, but constant fugacity. More specifically, Fig.\,\ref{fig:order_cubes} displays the radial and longitudinal order parameters of confined HCs at $f^{\star}=2000$, corresponding to $(\beta\mu^{ex},\rho T^3_p)=(11.043, 0.032)$, $(10.954, 0.035)$, and $(10.928, 0.036)$ at $R^{\star}=3\times\sqrt[3]{9}$, $6\times\sqrt[3]{9}$ and $9\times\sqrt[3]{9}$ respectively. In all cases,  $S^{\theta}_{2,\lambda_r} \approx S^{\theta}_{2,\lambda_l} \approx 0$ for $\lambda=$ $\hat{L}$, $\hat{W}$ or $\hat{T}$, suggesting a random orientation of HCs in both radial and longitudinal directions and across the concentric cylindrical shells. More interesting are the results reported in Fig.\,\ref{fig:ob_order_cubes}, where the order parameters $S^{\theta}_{2,\lambda_r}$ and $S^{\theta}_{2,\lambda_l}$ are shown for fluids of oblate HBPs. Also in this case, we display results at fugacity $f^{\star}=2000$, corresponding to values of excess chemical potential and density that are very similar to those of Fig.\,\ref{fig:order_cubes}. In particular, $(\beta\mu^{ex},\rho T^3_p)=(11.134, 0.029)$, $(11.098, 0.030)$, and $(11.078, 0.031)$ at $R^{\star}=9$, 12 and 15, respectively. Similar conclusions can also be drawn at other fugacities and are not discussed here. The spatial dependence of $S^{\theta}_{2,\lambda_r}$ and $S^{\theta}_{2,\lambda_l}$ indicates a significant orientation of oblate HBPs at the fluid/solid interface, where $\hat{T}$ is strongly aligned with the nanopore radius (empty triangles), whereas $\hat{W}$ and $\hat{L}$ are oriented perpendicularly to it (empty squares and circles), but not necessarily aligned along the nanopore axis (solid circles and squares). Such a particle anchoring on the wall is observed across the three nanopore radii studied here and even at relatively dilute systems, where $\rho T_p^3 \approx 10^{-2}$. At slightly larger distances from the walls, the  order parameter that measures the alignment of $\hat{T}$ with the nanopore radius decays to $S^{\theta}_{2,T_r} \approx -0.25$, suggesting that $\hat{T}$ is almost perpendicular to the nanopore radial direction (empty triangles). However, since $S^{\theta}_{2,T_l}$ is very low, the alignment between $\hat{T}$ and the nanopore longitudinal direction is basically random (solid triangle). In other words, at increasing distances from the wall, oblate HBPs gradually acquire a more and more disordered configuration that, in the nanopore core, eventually results in a fully isotropic one. Similarly, prolate HBPs orient at the fluid/solid interface and form nematic-like domains that are aligned along the longitudinal direction of the cylindrical nanopore. This is shown in Fig.\,\ref{fig:pr_order_cubes} for $f^{\star}=5000$, a value of fugacity that provides $(\beta\mu^{ex},\rho T_p^3)=(12.318, 0.022)$, $(12.275, 0.023)$, and $(12.219, 0.024)$ at $R^{\star}=12$, 15 and 22.5, respectively. In this case, the particle minor axes, $\hat{T}$ and $\hat{W}$, orient perpendicularly to the longitudinal direction of the cylindrical nanopore, with $S^{\theta}_{2,T_l} \approx S^{\theta}_{2,W_l} \approx -0.5$ (solid squares and triangles), while the particle main axis, $\hat{L}$, results to be parallel to it, with $S^{\theta}_{2,L_l} \approx 1$ (solid circles). As far as the radial alignment of the minor axes is concerned, we observe that $S^{\theta}_{2,T_r} \approx S^{\theta}_{2,W_r} \approx 0.25$ (empty triangles and squares), suggesting a very weak orientation at the walls that becomes completely random in the nanopore core.

\begin{figure}[!t]
\includegraphics[width=\columnwidth]{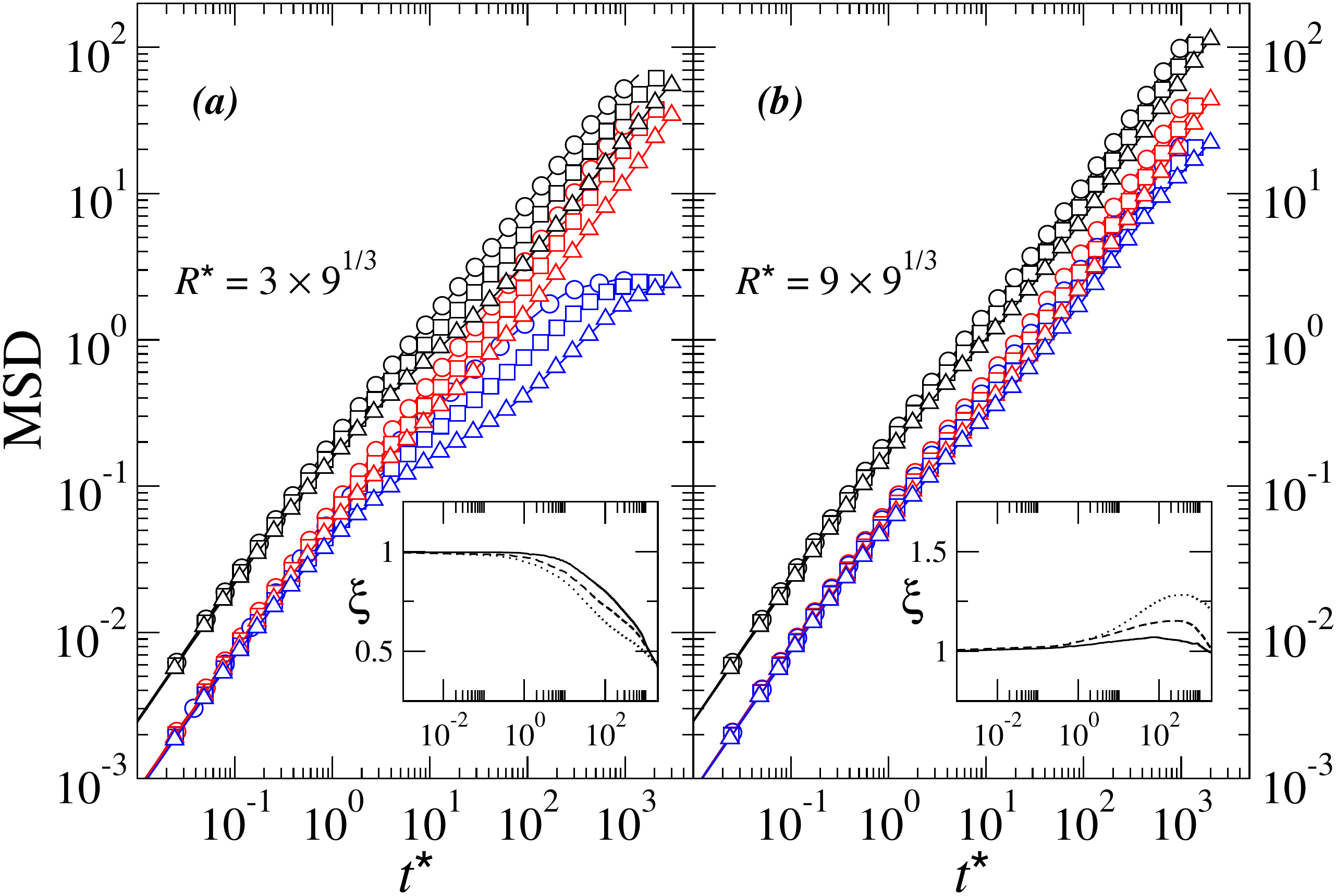}
\caption{Total (black symbols), longitudinal (red symbols) and radial (blue symbols) MSD of HCs confined in cylindrical nanopores of radius $R^{\star}=3\times\sqrt[3]{9}$ (\textit{a}) and $R^{\star}=9\times\sqrt[3]{9}$ (\textit{b}). Circles, squares and triangles refer, respectively, to $f^{\star}=5$, 50 and 500. The insets report the ratio $\xi(t)$ between the total MSD calculated in the nanopore and the total MSD calculated in the bulk at $f^{\star}=5$ (solid line), 50 (dashed line) and 500 (dotted line).}
\label{fig:MSD_cubes}
\end{figure}

Not only does confinement affect the phase behaviour and structural properties of HCs and HBPs, but it also determines their ability to diffuse through the nanopore. In order to estimate the effect of cylindrical confinement on the dynamics of HBPs, we have calculated their mean square displacement (MSD) in the direction parallel to the longitudinal axis of the nanopore and in the radial direction, and compared them with the total MSD. Longitudinal ($l$), radial ($r$), and total MSDs are defined as follows:

\begin{equation}
    \langle \Delta r^2(t) \rangle_{l}= \Bigg \langle \frac{1}{N} \sum_{j=1}^N [r_{j,l}(t) - r_{j,l}(0)]^2 \Bigg\rangle,
\end{equation}

\begin{equation}
    \langle \Delta r^2(t) \rangle_{r}= \Bigg\langle \frac{1}{N} \sum_{j=1}^N [r_{j,r}(t) - r_{j,r}(0)]^2 \Bigg\rangle,
\end{equation}

\begin{equation}
    \langle \Delta r^2(t) \rangle= \Bigg\langle \frac{1}{N} \sum_{j=1}^N [\textbf{r}_j(t) - \textbf{r}_j(0)]^2 \Bigg\rangle.
\end{equation}

\noindent with $r_{j,l}(t)$ and $r_{j,r}(t)$ the longitudinal and radial coordinates of particle $j$ at time $t$, respectively. The MSDs of HCs confined in nanopores of radius $R^{\star}=3\times\sqrt[3]{9}$ and $R^{\star}=9\times\sqrt[3]{9}$ are reported in Fig.\,\ref{fig:MSD_cubes} for different values of the excess chemical potential. Increasing $\mu^{ex}$ reduces the mobility of particles at long-time scales, where the effect of packing becomes especially relevant. By contrast, at short-time scales, when the particles are still very close to their original location, the MSD is practically unaltered by a change in chemical potential or, equivalently, density. While, at short times, the contribution of the radial and longitudinal MSDs to the total MSD is essentially indistinguishable, the effect of confinement appears to be evident at approximately $t^{\star}>1$, when the radial MSD slows down significantly as compared to the longitudinal MSD, especially so at large chemical potentials and narrow nanopore radii. In particular, the radial MSD in narrow nanopores exhibits a noticeable decrease that indicates the occurrence of a sub-diffusive dynamics that persists at long times. This is not the case in nanopores sufficiently large to allow HBPs to enter the diffusive regime.

\begin{figure}[!t]
\includegraphics[width=\columnwidth]{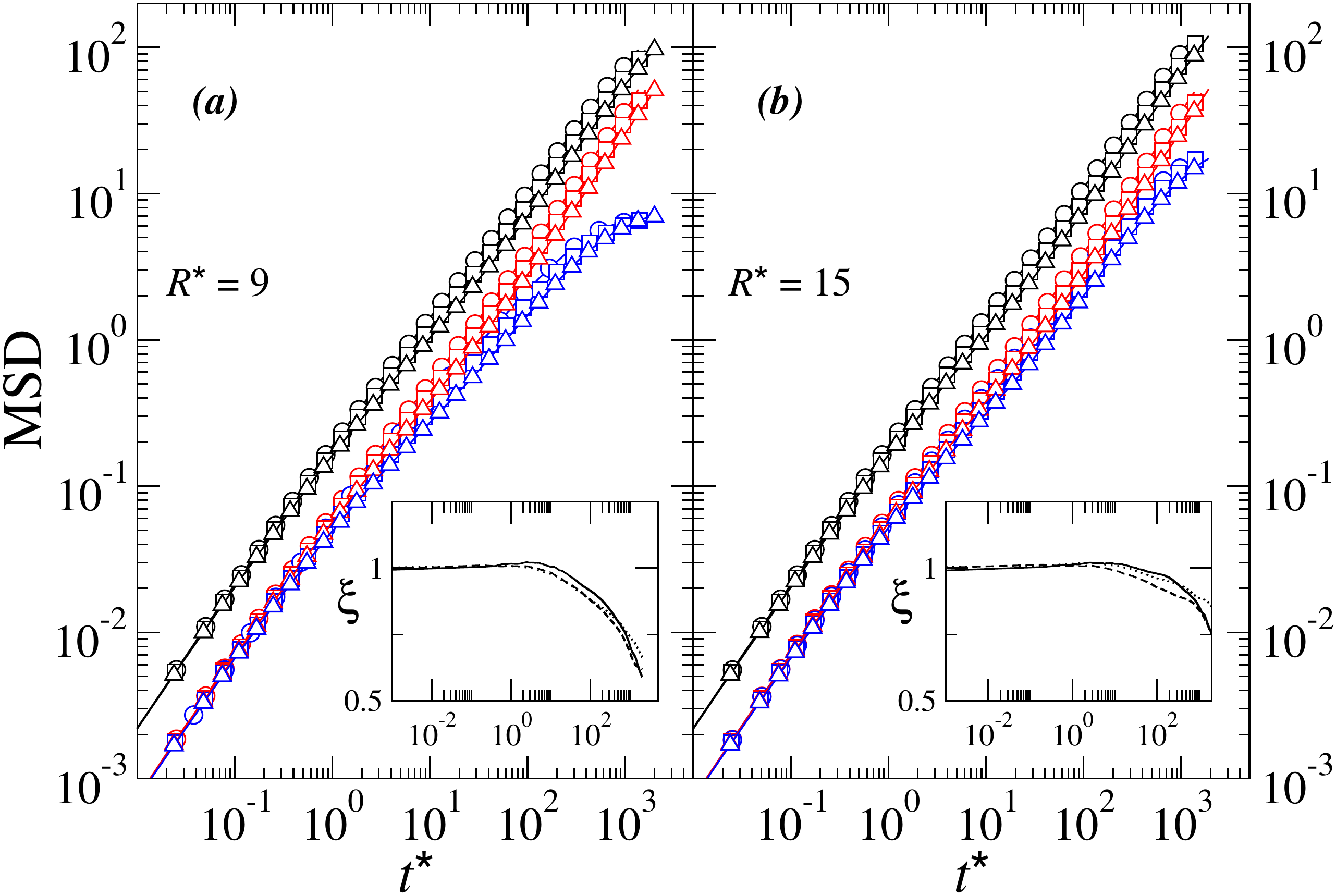}
\caption{Total (black symbols), longitudinal (red symbols) and radial (blue symbols) MSD of oblate HBPs confined in cylindrical nanopores of radius $R^{\star}=9$ (\textit{a}) and $R^{\star}=15$ (\textit{b}). Circles, squares and triangles refer, respectively, to $f^{\star}=5$, 50 and 500. The insets report the ratio $\xi(t)$ between the total MSD calculated in the nanopore and the total MSD calculated in the bulk at $f^{\star}=5$ (solid line), 50 (dashed line) and 500 (dotted line).}
\label{fig:MSD_ob_cub}
\end{figure}

\begin{figure}[!t]
\includegraphics[width=\columnwidth]{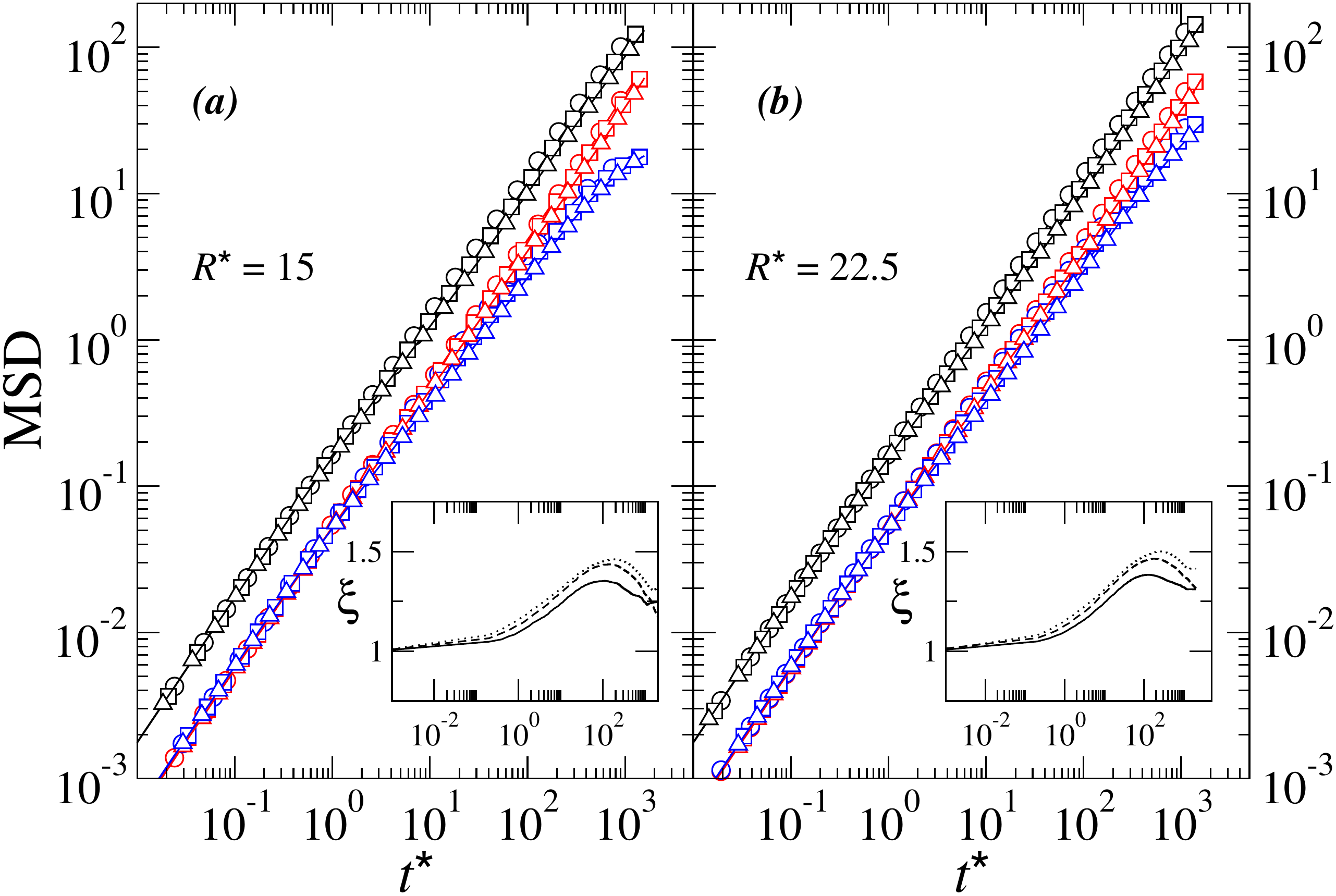}
\caption{Total (black symbols), longitudinal (red symbols) and radial (blue symbols) MSD of prolate HBPs confined in cylindrical nanopores of radius $R^{\star}=15$ (\textit{a}) and $R^{\star}=22.5$ (\textit{b}). Circles, squares and triangles refer, respectively, to $f^{\star}=5$, 50 and 500. The insets report the ratio $\xi(t)$ between the total MSD calculated in the nanopore and the total MSD calculated in the bulk at $f^{\star}=5$ (solid line), 50 (dashed line) and 500 (dotted line).}
\label{fig:MSD_pr_cub}
\end{figure}

The MSDs of oblate and prolate HBPs, reported respectively in Figs.\,\ref{fig:MSD_ob_cub} and \ref{fig:MSD_pr_cub}, exhibit qualitatively similar tendencies to those of the MSD of HCs, although some quantitative differences are noticed. At short-time scales, these differences, which depend exclusively on the diffusion coefficients at infinite dilution and hence only on particle geometry, are relatively small. Nevertheless, system packing, degree of confinement and long-range ordering become more and more relevant as time passes and eventually determine the ability of HBPs to diffuse in the nanopore. We stress that, although the MSDs in frames (\textit{a}) and (\textit{b}) of Figs.\,\ref{fig:MSD_cubes}-\ref{fig:MSD_pr_cub} have been calculated at the same chemical potential, the corresponding density changes with the nanopore radius, tending to the value in the bulk when the nanopore is sufficiently large (see Tables S1-S3 in the supplementary material). Interestingly, while denser suspensions would be expected to exhibit slower dynamics, their MSDs unambiguously indicate that the opposite behaviour is actually observed. This is especially evident in systems of HCs (Fig.\,\ref{fig:MSD_cubes}), where the total MSD at $t^{\star} \approx 10^3$ in nanopores of $R^{\star}=9\times \sqrt[3]{9}$ (frame (\textit{b})) is approximately 4 to 5 times larger than that in nanopores of $R^{\star}=3\times \sqrt[3]{9}$ (frame (\textit{a})), depending on fugacity. When one separately examines the two contributions to the total MSD, it turns out that the origin of this unexpected increase of mobility with packing is entirely due to the radial MSD, the longitudinal MSD being essentially independent on nanopore radius. We conclude that, in sufficiently narrow nanopores, the limitations imposed to radial diffusion dramatically determine the particle global mobility as much as, and even more than, system packing.

In the insets included in Figs.\,\ref{fig:MSD_cubes}-\ref{fig:MSD_pr_cub}, we report the ratio between the total MSD calculated in the nanopore and the total MSD calculated in the bulk, that is $\xi (t) \equiv \langle \Delta r^2(t) \rangle_{\rm pore}/\langle \Delta r^2(t) \rangle_{\rm bulk}$. At short time scales, when the dynamics is still insensitive to both packing and degree of confinement, the ratio between the nanopore and bulk MSDs displays the same general behaviour for all the systems studied, with $\xi=1$. However, at intermediate-to-long time scales, $\xi$ deviates from 1, suggesting that these factors along with particle geometry become relevant. The inset in the left frame of Fig.\,\ref{fig:MSD_cubes}, where $\xi<1$ for $t^{\star} > 1$, suggests that HCs diffuse faster in the bulk than in narrow nanopores. In larger nanopores (right frame of Fig.\,\ref{fig:MSD_cubes}), the behaviour is more complex, with $\xi$ increasing up to 1.25 at $f^{\star}=500$ and then decreasing at longer time scales, indicating that, at least over approximately three time decades, the dynamics under confinement is faster than the dynamics in the bulk. Beyond our simulation time, we would expect a similar behaviour to that observed in smaller nanopores, where $\xi<1$ at sufficiently long times. This view is further supported by the insets of Fig.\,\ref{fig:MSD_ob_cub}, which display $\xi$ for suspensions of oblate HBPs. Also in this case, $\xi$ decreases to values lower than unity when the long-time diffusive regime is achieved. A similar peak to that observed in the right inset of Fig.\,\ref{fig:MSD_cubes} (suspensions of HCs in nanopores of radius $R^{\star}=15$) is also observed in the insets of Fig.\,\ref{fig:MSD_pr_cub} (suspensions of prolate HBPs in nanopores of radius $R^{\star}=15$ and 22.5). The occurrence of such a maximum at intermediate time scales, when particles are still relatively close to their original position and about to diffuse through the \textit{cage} formed by their immediate neighbours, must be somehow connected to the topological differences existing between the cages in the bulk and those in a confined environment. While in an isotropic system, the time a particle takes to diffuse through a cage is not dependent on its direction in space, such a space invariancy does not hold upon constraining this system in a cylindrical nanopore. Confining the fluid induces relevant modifications of its structure that are not detected in the bulk, especially in the vicinity of the wall. The density profiles and uniaxial order parameters presented above confirm that this is indeed the case, and the difference between longitudinal and radial MSDs is an unequivocal evidence that through structure one can control dynamics. Nevertheless, the existence of a space-dependent cage effect does not necessarily result into a faster diffusion in a nanopore as compared to that in the bulk. For example, in suspensions of oblate HBPs, we find $\xi \leq 1$ in both narrow and large nanopores, from short to long time scales (see insets of Fig.\,\ref{fig:MSD_ob_cub}). While we believe that the morphology of the particle cage dramatically influences the dynamics at these time and length scales, further investigation is needed to fully understand what other elements contribute to enhance or inhibit the diffusion of a nanoconfined fluid as compared to its usual behaviour in the bulk. The presence of clusters, that we have recently detected in very similar systems of oblate and prolate HBPs, might contribute to answer this question \cite{tonti2021}. 

We now turn our attention to the distribution of the particle displacements over time. To this end, we calculated the self-part of the van Hove correlation function (s-VHF) in the longitudinal and radial directions of the cylindrical nanopore, which respectively read

\begin{equation}
    G_{s,l}(r_{l},t)=\frac{1}{N}\Bigg \langle \sum_{j=1}^{N}\delta(r_{l}-|{\bf{r}}_{j,l}(t)-{\bf{r}}_{j,l}(0)|) \Bigg \rangle
\end{equation}

and

\begin{equation}
    G_{s,r}(r_{r},t)=\frac{1}{N}\Bigg \langle \sum_{j=1}^{N}\delta(r_{r}-|{\bf{r}}_{j,r}(t)-{\bf{r}}_{j,r}(0)|) \Bigg \rangle.
\end{equation}

\begin{figure}[t!]
\includegraphics[width=\columnwidth]{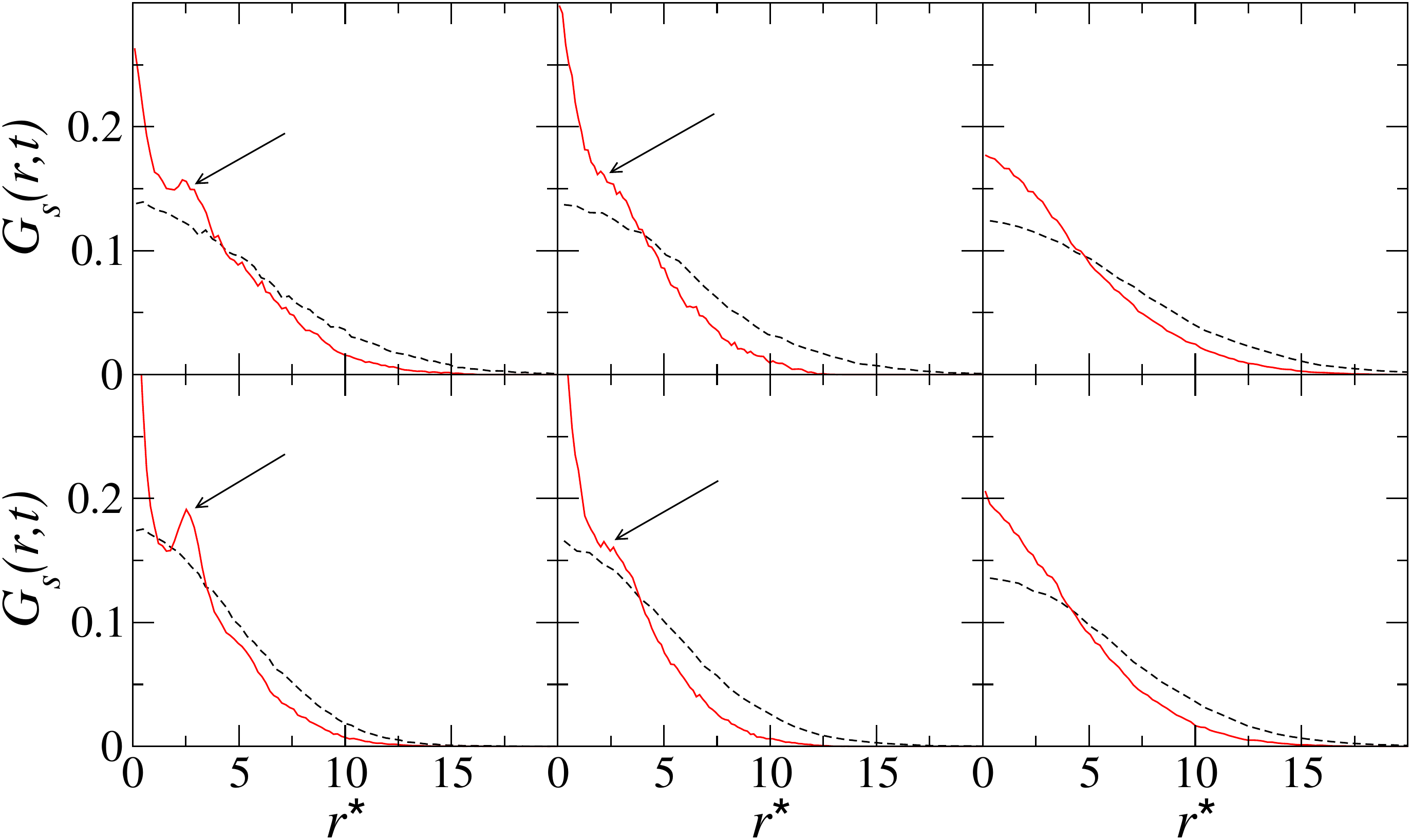}
\caption{Longitudinal (black dashed line) and radial (solid red line) s-VHFs obtained at $t^{\star} \approx 10^3$ in suspensions of HCs (left), oblate HBPs (middle) and prolate HBPs (right) confined in cylindrical nanopores of radius $R^{\star}=9\times \sqrt[3]{9}$, 15 and 22.5, respectively. Top and bottom frames refer to $f^{\star}=5$ and $f^{\star}=500$, respectively. The arrows indicate the occurrence of a pick or shoulder observed in the radial s-VHFs of HCs and oblate HBPs, respectively.}
\label{fig:VHFs}
\end{figure}

\noindent In particular, the symbol $\delta$ is the Dirac delta and the angular brackets denote ensemble average over the $N$ particles and at least 100 uncorrelated DMC trajectories. Longitudinal and radial s-VHFs have been normalised as follows: $\int_{0}^{\infty}G_{s,l} dr_{l} = \int_{0}^{\infty} G_{s,r} dr_{r}$=1. The s-VHFs are especially relevant to estimate the probability of observing particles that are slower or faster than the average. Considering the tendencies uncovered by radial and longitudinal MSDs, these probabilities are expected to be different along the nanopore axis and in the radial direction. This is indeed the case as one can appreciate from Fig.\,\ref{fig:VHFs}, where radial and longitudinal s-VHFs of the same systems as those considered in  Figs.\,\ref{fig:MSD_cubes}-\ref{fig:MSD_pr_cub} are reported. More specifically, we present the radial (solid lines) and longitudinal (dashed lines) s-VHFs of HCs and HBPs at $f^{\star}=5$ (top frames) and 500 (bottom frames) and at time $t^{\star} \approx 10^3$. While the two sets of s-VHFs appear relatively similar, thus suggesting quasi-isotropic particle displacements in the nanopore, there are distinctive features that deserve further attention. In particular, the longitudinal s-VHFs exhibit a peak at small distances that appears slightly more pronounced at increasing fugacity, where the system density is larger and it is more difficult for the particles to diffuse. This behaviour is especially evident in systems of HCs and oblate HBPs, while the s-VHFs of prolate HBPs display a weaker dependence on fugacity. For the three particle geometries, the longitudinal s-VHF gradually vanishes with the distance, suggesting that the probability of finding \textit{fast} particles, namely particles that are able to displace longer distances over the same time window, is smaller and smaller and eventually decays to zero. As far as the radial s-VHFs are concerned, the very high peak at $r^{\star} \approx 0$ suggests that, even at relatively long time scales, the mobility of particles across the nanopore concentric shells is quite limited. Additionally, the analysis of these s-VHFs discloses a few interesting elements that are strongly related to the density profiles of Fig.\,\ref{fig:dens_prof}. 

In particular, the radial s-VHFs of confined HCs (left frames in Fig.\,\ref{fig:VHFs}) exhibit a minor peak between $r^{\star}=3$ and 4. As the left frame of Fig.\,\ref{fig:dens_prof} indicates, the fluid in the cylindrical nanopore does not exhibit a homogeneous density distribution in the vicinity of the pore wall, but organises in concentric shells intercalated by domains that are less likely to be occupied. Correspondingly, particles displacing back and forth between these contiguous shells are less likely to linger in between them, producing a probability of displacements that, at sufficiently large densities, results in the minor peaks observed in the left frames of Fig.\,\ref{fig:VHFs}. Similar considerations are still valid for the radial s-VHFs of oblate HBPs (middle frames of Fig.\,\ref{fig:VHFs}). However, in this case the minor peaks are much less pronounced than those observed for HCs, almost degenerating into a smooth shoulder. This is consistent with the corresponding density profiles of Fig.\,\ref{fig:dens_prof} (middle frames), which indicate a milder density differences between the two outermost cylindrical shells. By contrast, no secondary peaks are observed in the radial s-VHFs of prolate HBPs (right frames of Fig.\,\ref{fig:VHFs}), in agreement with the particle density distribution of Fig.\,\ref{fig:dens_prof}, which exclude the existence of well-defined shells at the nanopore wall as observed for HCs and oblate HBPs. We finally notice that the radial s-VHFs gradually decrease to zero at distances approximately between $r^{\star}=15$ and 20. Such a decrease indicates a non-vanishing probability to observe especially fast particles that have displaced significantly longer distances than the average particle. The simultaneous presence or slow and fast particles is also detected in the longitudinal s-VHFs (dashed lines in Fig.\,11), although in this case their decrease is monotonic with no evidence of peaks or shoulders.


\section{CONCLUSIONS}
In summary, we have here analysed the behaviour of colloidal hard cubes and hard cuboids that have been confined in cylindrical nanopores. We have specifically assessed the effect of packing and particle and pore geometry on the system equilibrium and dynamical properties, finding that both factors directly determine the ability of HCs and HBPs to orient at the fluid/solid interface and their mobility along the nanopore longitudinal and radial axes. In particular, oblate and prolate HBPs, at sufficiently large density, organise in ordered nematic-like shells that surround an inner core of randomly-oriented particles. This is not the case for HCs, which maintain a random orientation throughout the cylindrical nanopore. Regardless the occurrence of nematic-like ordering, the radial density profiles, obtained along the nanopore radial direction, indicate the existence of a layered arrangement that is especially evident in systems of HCs, but less so in systems of prolate HBPs. These structural signatures dramatically determine the dynamical properties of the systems under confinement, which we have evaluated in terms of MSD and s-VHFs. The computation of the MSD is very useful to analyse the mobility of HBPs under confinement as compared to that in the bulk. We have shown that at short-time scales, HBPs diffuse in the nanopore as they were in the bulk. Then, at intermediate times, the combined action of layering and alignment in confined suspensions enhances the ability of particles to diffuse through their surrounding neighbours, leading to a dynamics that is faster than that in the bulk. However, these tendencies are only limited to a few time decades. As soon as the system enters the long-time diffusive regime, the ratio between the MSD in the nanopore and that in the bulk gradually decreases, eventually approaching values lower than one beyond our simulation time. Finally, the s-VHFs have been key to distinguish the distribution of particle displacements along the nanopore longitudinal and radial directions. Both of them suggest the existence of particles that displace shorter or longer distances than the average particle. In particular, radial s-VHFs are especially interesting as they reveal the simultaneous presence of slow particles that remain close to their original position even at relatively long times, and fast particles that are able to cover distances as large as 15 to 20 times that covered by most particles over the same time scale. A similar tendency, although on a minor scale, is also found along the longitudinal direction of the nanopore.


\begin{acknowledgments}
A.P. acknowledges financial support from the Leverhulme Trust Research Project Grant No.\,RPG-2018-415 and the use of the Computational Shared Facility at the University of Manchester. A.C. acknowledges the Spanish Ministerio de Ciencia, Innovación y Universidades and FEDER for funding (Project No.\,PGC2018-097151-B-I00) and C3UPO for the HPC facilities provided.
\end{acknowledgments}

\section{Data Availability}
The data that support the findings of this study are available from the corresponding author upon reasonable request.

\nocite{*}
\bibliography{aipsamp}

\end{document}